\documentclass{article}
\usepackage{ijcai19}

\usepackage{times}
\usepackage{soul}
\usepackage{url}
\usepackage[hidelinks]{hyperref}
\usepackage[utf8]{inputenc}
\usepackage[small]{caption}
\usepackage{graphicx}
\usepackage{amsmath}
\usepackage{booktabs}
\usepackage[skip=0pt]{caption}
\usepackage{enumitem}
\setenumerate[1]{itemsep=0pt,partopsep=0pt,parsep=\parskip,topsep=5pt}
\setitemize[1]{itemsep=0pt,partopsep=0pt,parsep=\parskip,topsep=5pt}
\setdescription{itemsep=0pt,partopsep=0pt,parsep=\parskip,topsep=5pt}

\usepackage[sort&compress,square,comma,authoryear]{natbib} 

\usepackage{color}
\usepackage{amsmath}
\usepackage{amsfonts}
\usepackage{latexsym}
\usepackage{acronym}

\acrodef{TDS}{Task-oriented Dialogue System}
\acrodef{MemNN}{End-to-End trainable Memory Network}
\acrodef{EntNet}{Recurrent Entity Network}
\acrodef{SEntNet}{Source-aware Recurrent Entity Network}
\acrodef{RNN}{Recurrent Neural Network}
\acrodef{POS-EntNet}{Recurrent Entity Network Enriched with Part-of-speech Tags}
\acrodef{POS-SEntNet}{Source-aware Recurrent Entity Network enriched with part-of-speech tags}
\acrodef{DRS}{Dialogue Response Selection}
\acrodef{NLU}{Natural Language Understanding}
\acrodef{DST}{Dialogue State Tracking}
\acrodef{PL}{Policy Learning}
\acrodef{NLG}{Natural Language Generation}
\acrodef{POS}{Part-Of-Speech}
\acrodef{KB}{knowledge base}
\acrodef{NUC}{Next-Utterance-Classification}
\acrodef{BDC}{bAbI Dialog Challenge}

\newcommand{\Source}{\mathcal{S}}
\newcommand{\SourceUser}{\mathcal{S_{\mathbb{U}}}}
\newcommand{\SourceSystem}{\mathcal{S_{\mathbb{S}}}}
\newcommand{\SourceKB}{\mathcal{S_{\mathbb{B}}}}

\newcommand{\OurParagraph}[1]{%
\smallskip\noindent\textbf{#1}~%
}

\newcommand*\titleheader[1]{\gdef\@titleheader{#1}}

\usepackage{fancyhdr}
\title{SEntNet: Source-aware Recurrent Entity Network\\ for Dialogue Response Selection}

\author{
Jiahuan Pei$^1$\and
Arent Stienstra$^1$\and
Julia Kiseleva$^2$\And
Maarten de Rijke$^1$\\
\affiliations
$^1$University of Amsterdam\\
$^2$Microsoft Research AI\\
\emails{
\{j.pei, derijke\}@uva.nl,
arent.stienstra@gmail.com, julia.kiseleva@microsoft.com}
}

\begin{document}

\maketitle

\begin{abstract}
\emph{Dialogue response selection} is an important part of \acp{TDS}; it aims to predict an appropriate response given a dialogue context.
Obtaining key information from a complex, long dialogue context is challenging, especially when different sources of information are available, e.g., the user's utterances, the system's responses, and results retrieved from a \ac{KB}.
Previous work ignores the type of information source and merges sources for response selection.
However, accounting for the source type may lead to remarkable differences in the quality of response selection. 
We propose the \acf{SEntNet}, which is aware of different information sources for the response selection process.
\ac{SEntNet} achieves this by employing source-specific memories to exploit differences in the usage of words and syntactic structure from different information sources (user, system, and \ac{KB}). 
Experimental results show that \ac{SEntNet} obtains 91.0\% accuracy on the Dialog bAbI dataset, outperforming prior work by 4.7\%. 
On the DSTC2 dataset, \ac{SEntNet} obtains an accuracy of 41.2\%, beating source \emph{unaware} recurrent entity networks by 2.4\%.
\end{abstract}


\section{Introduction}
\label{sec:introduction}

\begin{figure}[t]
    \centering
    \includegraphics[width=\columnwidth]{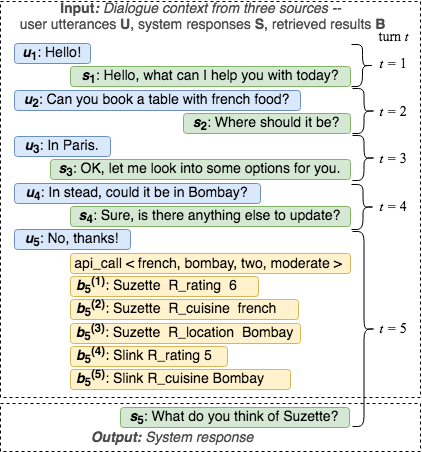}
    \caption{An example of response selection for booking a restaurant. The top box contains the input for response selection; the bottom box shows the selected response.}
    \label{fig:example_dialogue}
\end{figure}



\acfp{TDS} have attracted a lot of attention recently for their practical applications, e.g., booking flight tickets or scheduling meetings~\citep{young2013pomdp,williams2017hybrid}.
Unlike open-ended dialogue systems~\citep{zhao2016towards}, \acp{TDS} aim to assist users to achieve specific goals through multiple dialogue turns.

Recently introduced end-to-end approaches for \acp{TDS} improve over traditional ones~\citep{end2end_dataset_paper_babi_bordes,survey_dialog_systems,young2013pomdp} due to their ability to deal with global optimization, which facilitates easier adaptation to new domains~\citep{survey_dialog_systems}. 
End-to-end approaches can be classified into two categories: \textit{response generation} and \textit{response selection}. 
We focus on response selection methods because they show more convincing performance than response generation ones~\citep{eric2017key,wen2016network}. 
Different information sources may contribute to the response selection process.
It is known that the combination of specialized models may be more effective than a single general model~\citep{dietterich2000ensemble}.
Thus, we hypothesize that the combination of source-specific expert models outperforms a single general model in a response selection process.
Let us consider an example of dialogue response selection for a five-turn dialogue shown in Figure~\ref{fig:example_dialogue}.
In turn 5 ($t=5$), the system should give a recommendation in response to the user utterance ``No, thanks!" 
This recommendation has to include an entity, i.e., the name of a restaurant (``Suzette''), that has not occurred yet in the dialogue history and, hence, has to come from the \acf{KB} that feeds the dialogue.
So, a model that decides which response to select should prefer to consider the \ac{KB} rather than the user utterances or past system responses so as to obtain the entity (``Suzette'') for its response.
Awareness of the source of information is essential for dialogue language understanding~\citep{chen2017dynamic}. 
However, state-of-the-art response selection methods~\citep{end2end_dataset_paper_babi_bordes,liu2017gated,dstc6_memnet_contextnr,wu2018end} 
are not explicitly aware of the different information sources that play a role in response selection.

We propose an end-to-end response selection model, named \acfi{SEntNet} that is source-aware.
As an extension of \acs{EntNet}~\citep{entnet}, \ac{SEntNet}  provides dynamic long-term memory blocks to maintain and update latent concepts and attributes. 
\ac{SEntNet} is designed to improve \acs{EntNet} by introducing source-specific memories to exploit differences in the usage of words and syntactic structure in different information sources.
Moreover, \ac{SEntNet} employs a source-aware attention mechanism to dynamically capture the importance of different sources.
\ac{SEntNet} is able to choose responses more effectively by exploiting source-specific features derived from the dialogue history and \ac{KB}.
Furthermore, the computational costs implied by the use of extra memory modules can be offset by a parallel update mechanism design.

Our main contribution is the \ac{SEntNet} model that significantly outperforms \acs{EntNet} in terms of the turn-level accuracy of response selection. 
We carry out extensive experiments on the Dialog bAbI and modified DSTC2 datasets.
Our experimental analysis shows the following properties of \ac{SEntNet}: 
\begin{itemize}[leftmargin=*,nosep]
    \item an ability to capture the semantics of dialogue context and learn word embeddings during the training process; 
    \item a tolerance against sparse data, that is, it displays a stable performance, even when trained on a small amount of training data; and
    \item an ability to handle different degrees of lexical diversity, which may be affected by noise.
\end{itemize}

\section{Problem Definition} 
\label{sec:problem_formulation}

\ac{TDS} can be framed as a collection of search constraints that are denoted by \textit{slots} $\{k_1, \ldots, k_l\}$ and \textit{values} $\{v_1, \ldots, v_l\}$ that each slot can take. 
Result snippets can be obtained by issuing a symbolic query $api\_call=(v_1, \ldots, v_l)$ to a \ac{KB} that is generated by those search constraints.
Each retrieved result is an entry from the \ac{KB} that can be presented as a triple (\textit{entity$_1$}, \textit{relation}, \textit{entity$_2$}).
E.g., in the restaurant booking domain, such a triple could look like (\textit{Noma}, \textit{cuisine-type}, \textit{Indian}) and the relations include \emph{rating}, \emph{cuisine-type} and \emph{location}, etc.

The \emph{dialogue context} is used for response selection.
It consists of alternating utterances with three main \emph{sources}, i.e., the \emph{user}, the \emph{system} and \emph{retrieved results} from the \ac{KB}.
Formally, the dialogue context at \textit{turn} $t$ is defined as a tuple $(\mathbb{U}_t, \mathbb{S}_{t-1}, \mathbb{B}_t)$ where:
\begin{itemize}[leftmargin=*,nosep]
\item $\mathbb{U}_t=(u_1$, $u_2$, \ldots, $u_t)$ are user utterances, which are highlighted in blue in Figure~\ref{fig:example_dialogue};
\item $\mathbb{S}_{t-1}=(s_1$, $s_2$, \ldots, $s_{t-1})$ are system responses, which are highlighted in green in Figure~\ref{fig:example_dialogue}; and
\item $\mathbb{B}_{t}=(b^1_t, b^2_t,\ldots, b^\lambda_t)$ is a sequence of $\lambda$-best retrieved results from an external \ac{KB}, which are highlighted in yellow in Figure~\ref{fig:example_dialogue}. 
\end{itemize}
Therefore, we consider three sources of information $\Source \in \{\SourceUser, \SourceSystem, \SourceKB\}$ for response selection.

Figure~\ref{fig:example_dialogue} shows an example of a dialogue turn where the system needs to select a restaurant suggestion satisfying the user's needs. 
In this case, only two sources are useful: $\mathbb{U}$ because it contains user preferences and $\mathbb{B}$  because it has information about available restaurants.
%

We aim to learn a response selection model $\psi$, parameterized by $\Theta$, that predicts a candidate response $s_{t}$ by taking as input a dialogue context $(\mathbb{U}_t, \mathbb{S}_{t-1}, \mathbb{B}_t)$ and is able to decide which sources of information are most useful at turn $t$:
\begin{equation}
\textstyle
\label{eq:dialogue_system_equation}
    \psi_\Theta(\mathbb{U}_t, \mathbb{S}_{t-1}, \mathbb{B}_t) \xrightarrow{} s_{t}.
\end{equation}


\section{Preliminaries: Recurrent Entity Networks}
\label{sec:background}

Unlike Memory Networks~\citep{memory_networks}, \acs{EntNet} is able to manage entities that are contained in \ac{KB} triples to track the state of the dialogue. 
\acs{EntNet} uses an attention mechanism in combination with \acp{RNN} to store and retrieve memories in parallel rather than sequentially.
\acs{EntNet}'s functions depend on three modules described below.

\OurParagraph{Input module.}
An utterance representation is obtained by multiplying the embedding vector of constituent words with a positional mask and then averaging the results.
Each utterance has an index $i$ to present its temporal position in the sequential dialogue history; $e_i  \in \mathbb{R}^d $ represents the embedding vector of the $i$-{th} utterance.

Let ${w}_{x}^{i} \in \mathbb{R}^d$ be the embedding vector of the $x$-{th} word of the $i$-{th} utterance, where the hyper-parameter $d$ is the dimension of the embedding vector.
The learnable parameter ${f}_x \in \mathbb{R}^d$ is the mask that is multiplied with the word embedding vector at position $x$.
The embedding vector of the $i$-{th} utterance is then calculated as:
\begin{equation}
\textstyle
    {e}_i = \sum_{x} {f}_x \odot {w}_{x}^{i} \in \mathbb{R}^d.
\end{equation}

\OurParagraph{Dynamic memory module.}
The state of entities is learned via memory blocks, each of which is a gated \acs{RNN} that encodes the information of one entity. 
The memory module has $m$ memory blocks and every block learns an embedding vector of the dialogue history with $n$ utterances. 
The $j$-{th} memory block of the $i$-{th} utterance has an embedding vector of a slot ${k_j^i} \in \mathbb{R}^d$ and a hidden state ${h_j^i} \in \mathbb{R}^d$.  
The gate $g_j^i \in \mathbb{R}^d$ determines how much information from the $i$-{th} utterance should influence the state of the $j$-{th} memory.
The learnable matrices $G \in \mathbb{R}^{d \times d}$, $V \in \mathbb{R}^{d \times d}$ and $W \in \mathbb{R}^{d\times d}$ are shared among different blocks;
$\odot$ denotes the element-wise product;
$\sigma$ and $\phi$ denote the sigmoid function and ReLU function, respectively.
For the $i$-{th} utterance in a dialogue, the memory block of the $j$-{th} entity is updated by:
\begin{align}
    g_j^i &= \sigma(e_i^T h_j^{i-1}  +e_i^T k_j^{i-1} ) \in \mathbb{R}^d \label{eq4}\\
    \tilde{h}_j^i &= \phi(G h_j^{i-1} + V{k_j}^{i-1} + W{e_i}) \in \mathbb{R}^d \label{eq5}\\
    h_j^i &= \dfrac{h_j^{i-1} + g_j^i \odot \tilde{h}_j^i}{\| h_j^{i-1} + g_j^i \odot \tilde{h}_j^i \| } \in \mathbb{R}^d \label{eq6}.
\end{align}
The final hidden state $h_j$ of the $j$-{th} memory block is the concatenation of the hidden states $\{h_j^1$, $h_j^2$, \ldots, $h_j^n\}$. 
The state for each memory block is initialized with the slot, i.e., $h_j^i = k_j^i$. 

\begin{figure*}[t]
    \centering
    \includegraphics[width=0.9\textwidth]{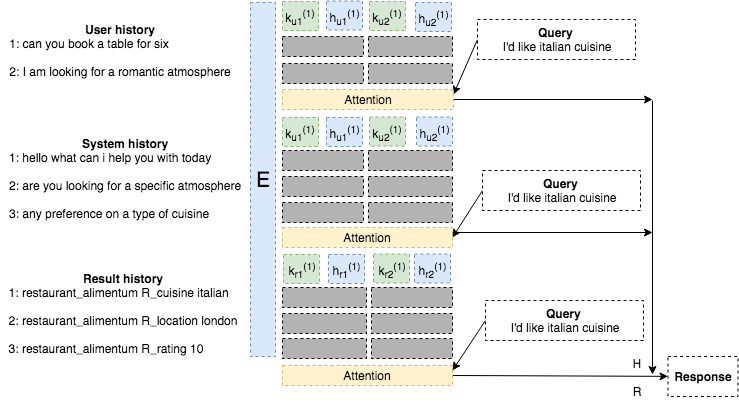}
    \caption{A schematic representation of the \ac{SEntNet} architecture with separate source-specific memory modules.}
    \label{fig:separate_input_sources}
\end{figure*}

\OurParagraph{Output module.}
\label{entnet_output_model}
The weighted sum of the final states $\{h_1, h_2, \ldots, h_m\}$ is used to predict the output, where the attention between those states and the representation of the query is used as the weights.
Let $q\in \mathbb{R}^d$ be the embedding vector of the user utterance $u_t$ for the current turn $t$. 
Also, $z \in \mathbb{R}^d$ is the weighted sum over the previous final states, where $p_j$ is the weight for the $j$-{th} state. 
The learnable matrix $L \in \mathbb{R}^{c \times d}$ is used to map the intermediate representation to the output distribution, where $c$ is the number of the candidate responses.
The weight matrix $H \in \mathbb{R}^{d \times d} $ is a learnable parameter. 
The intermediate prediction $\tilde{y} \in \mathbb{R}^{c}$ is a weighted sum between $q$ and $z$, normalized by the ReLU function $\phi$. 
The output $y \in \mathbb{R}^{c}$ is a distribution over the candidate responses, which can be calculated as follows: 
\begin{align}
    p_j &= \textstyle\dfrac{\exp{(q^T h_j)}}{ \sum_{k} \exp{(q^T h_k)}} \in \mathbb{R}^1 \\
    z &= \textstyle\sum_{j} h_j p_j \in \mathbb{R}^d \\ 
    \tilde{y} &=  L \phi(q + Hz) \in \mathbb{R}^c \\
    y &= \textstyle\dfrac{\exp{(\tilde{y}_j)}}{ \sum_{k} \exp({\tilde{y}_k)}} \in \mathbb{R}^c.
\end{align}


\section{Source-Aware Recurrent Entity Networks}
\label{sec:methods}

We propose the \acf{SEntNet}, which aims to learn an intermediate representation for each source of information with a separately parameterized memory module; see Figure~\ref{fig:separate_input_sources}. 
The source-specific memory modules share the embedding matrix $E$. 
The final state of every memory module is a weighted sum of the value vectors of the blocks in the module. 
The weights are calculated with attention between the query and the slots of the memory blocks. 
The output is then calculated by concatenating the attention weighted states of each memory module.

\OurParagraph{Input module.} 
For a specific information source $\Source$, each utterance has a new index $i$ to represent its temporal position in the sequential history originating from the source $\Source$. 
Let ${w}_{x}^{i}$ be the word embedding and ${l}_{x}^{i}$ be an optional item -- the \ac{POS} tag embedding of the $x$-{th} word in the $i$-{th} utterance of the source $\Source$; $f_x \in \mathbb{R}^d $ is the mask that is multiplied with the word embedding at position $x$. 
The embedding of the $i$-{th} utterance $e_{i(\Source)}$ for source $\Source$ is: 
\begin{equation}
\textstyle
    e_{i(\Source)} = {\sum_x} f_x \odot {w}_{x}^{i} +{l}_{x}^{i} \in \mathbb{R}^d.
\end{equation}

\OurParagraph{Dynamic memory module.} 
For the $i$-{th} utterance with a specific information source $\Source$ in the dialogue, the memory block for the $j$-{th} entity is updated as the following equations:
\begin{align}
    g_{j(\Source)}^i &= \sigma(e_{i(\Source)}^T h_{j(\Source)}^{i-1} + e_{i(\Source)}^T k_{j(\Source)}^{i-1}) \in \mathbb{R}^{d} \\
    \tilde{h}_{j(\Source)}^i &= \phi(G_{\Source} h_{j(\Source)}^{i-1} + V_{\Source} k_{j(\Source)}^{i-1} + W_{\Source} e_{i(\Source)})  \in \mathbb{R}^{d} \\
    h_{j(\Source)}^i &= \dfrac{h_{j(\Source)}^{i-1} + g_{j(\Source)}^i \odot {\tilde{h}_{j(\Source)}}^i}{\|h_{j(\Source)}^{i-1} + g_{j(\Source)}^i \odot {\tilde{h}_{j(\Source)}}^i \|} \in \mathbb{R}^{d}.
\end{align}
The final hidden state $h_{j(\Source)}$ of the $j$-{th} memory block is the concatenation of the hidden states $\{h_{j(\Source)}^1, h_{j(\Source)}^2, \ldots, h_{j(\Source)}^n\}$. 
The gate $g_{j(\Source)}^i$ controls how much the $i$-{th} utterance of source $\Source$ should contribute to the content of the $j$-{th} memory. 
$U_{\Source}$, $V_{\Source}$ and $W_{\Source}$ are trainable weight matrices that are separately parameterized among all the blocks. 

\OurParagraph{Output module.} 
Let $q\in \mathbb{R}^d$ be the embedding of the user utterance $u_t$ for the current turn $t$. 
The output module is defined as:
\begin{align}
    p_{j(\Source)} &= \textstyle\dfrac{\exp{(q^T h_{j(\Source)})}}{ \sum_{k} \exp({q^T h_{k(\Source)})}} \in \mathbb{R}^1 \\
    z_{\Source} &= \textstyle\sum_{j} h_{j(\Source)} p_{j(\Source)} \in \mathbb{R}^{d} \\ 
    z &= concat(z_{\SourceUser}, z_{\SourceSystem}, z_{\SourceKB}) \in \mathbb{R}^{3d} \\
    \tilde{y} &=  L \phi(q + Hz) \in \mathbb{R}^r \\
    y &= \textstyle\dfrac{\exp{(\tilde{y}_j)}}{ \sum_{k} \exp({\tilde{y}_k)}} \in \mathbb{R}^r.
\end{align}
Unlike \acs{EntNet}, an attention mechanism is used between the query and the memory blocks of each source-specific memory module, which is denoted as $p_{j(\Source)}$. 
It is helpful to think of $z_{\SourceUser}$, $z_{\SourceSystem}$, $z_{\SourceKB}$ as the outputs of expert models, each of which is specialized for a single information source. 
A policy is adopted to learn the final decision based on the prediction of each of the experts.
Here, we use concatenation to produce a single output vector $z$. 
The dimensionality of the learnable matrix $H$ is changed to $\mathbb{R}^{d \times 3d}$.

To sum up, \ac{SEntNet} enhances \acs{EntNet} by taking advantage of specialized predictions of ``experts'' that are modelled as source-specific memory modules to capture different sources of information -- so as to be able to select more appropriate responses given a user's intent.

\section{Experimental Setup}
\label{sec:experimental_setup}

We aim to answer the following research questions:
\textbf{RQ1:}~How well does \ac{SEntNet} predict appropriate responses?
\textbf{RQ2:}~How do different embeddings affect \ac{SEntNet}'s performance?
\textbf{RQ3:}~How well does \ac{SEntNet} perform in the case of limited data? And
\textbf{RQ4:}~How does lexical diversity affect \ac{SEntNet}'s performance?

\OurParagraph{Datasets and Evaluation.}
\label{sec:experiments_dataset}
We evaluate \ac{SEntNet} and other response selection models on two datasets: \textbf{dialog bAbI}~\citep{end2end_dataset_paper_babi_bordes} and \textbf{mDSTC2} -- a modified version of the DSTC2~\citep{henderson2014second}. 
Supporting facts from an external \ac{KB} are incorporated into the dialogue history. 
Out-of-dialogue information can be taken into account in the response selection process, e.g., when the list of restaurants retrieved from the \ac{KB} are considered together with the dialogue history, the response selection model can use these results to recommend a restaurant.

The dialog bAbI dataset consists of 3,000 noise-free simulated dialogues with 3,747 unique words and 4,212 candidate responses. 
We split it equally for training, validation, and testing. 
The mDSTC2 dataset consists of 2,785 real human-machine dialogues with 1,229 unique words and 2,406 candidate responses. 
It is divided into 1,168/500/1,117 dialogues for training, validation, and testing, respectively. 

As evaluation metric we use \textit{turn-level accuracy}, which is defined as the fraction of correct responses out of all responses. 
We utilized a paired t-test to show statistical significance ($p < 0.01$) of the relative improvements. 

\OurParagraph{Experiments.}
\label{sec:experiments}
%
\noindent
\textbf{(E1)} To answer \textbf{RQ1}, we evaluate the turn-level accuracy of \ac{SEntNet} against the following baselines:
 \begin{itemize}[leftmargin=*,nosep]
   \item \textbf{TF-IDF}. This model ranks candidate responses by TF-IDF weighted cosine similarity between one-hot vectors of input and candidate responses.
   \item \textbf{Query-to-answer (Q2A)}. Given a query, it finds the most common response in the train set~\citep{babi}. 
   \item \textbf{DQMemNN}. This is the state-of-the-art for response selection on dialog bAbI dataset~\citep{wu2018end}; 
   for a fair comparison, we used DQMemNN without exact matching and delexicalization.   
   \item \textbf{HHCN}. This is the state-of-the-art for response selection on the DSTC2 dataset~\citep{liang2018hierarchical}.
   \item \textbf{\acs{EntNet}}. We reproduced \acs{EntNet}, which was originally introduced for question answering and is reported to have strong reasoning abilities~\citep{entnet}. 
 \end{itemize}

\noindent
\textbf{(E2)} To answer \textbf{RQ2}, we compare the \textit{turn-level accuracy} of \ac{SEntNet} with different  pre-trained embeddings, i.e., Glove~\citep{glove}, Paragram~\citep{paragram} and NumberBatch~\citep{numberbatch}, with three embedding initialization strategies:
\begin{enumerate}[leftmargin=*,nosep]
    \item \textbf{Random strategy.} The word embeddings are initialized by a zero-mean Gaussian distribution with a standard deviation of $0.1$. They are optimized during training. 
    \item \textbf{Fixed strategy.}  This initialization method is similar to the one above but the embeddings are not optimized during training. It tests whether the model requires an embedding space that encodes the semantics between words.
    \item \textbf{Oracle strategy.} The model is trained on the full training set using the \textit{Random strategy}.
    We then obtain oracle embeddings that are optimized by all available training data. 
\end{enumerate}

\noindent
\textbf{(E3)} To answer \textbf{RQ3}, we train \ac{SEntNet} based on various fractions of the datasets and test how \ac{SEntNet} performs compared to the baselines even when limited training data is available.

\noindent
\textbf{(E4)} To answer \textbf{RQ4}, we simulate the case when \acs{SEntNet} is trained on a dataset with low lexical diversity. 
Specifically, we use SpaCy\footnote{https://spacy.io/usage/linguistic-features} to get the \ac{POS} tag of each word. Then the embeddings a tag is added to that of the word and can be optimized in the learning process using the random strategy. 

\OurParagraph{Training details.}
The training loss is measured by the log of the cross-entropy between the one-hot encoded golden label and the predicted output. 
Our objective is to minimize the loss over all of the $n$ samples with $c$ response candidates.
All weights of the network are initialized with a zero-mean Gaussian distribution using a standard deviation of $0.1$. 
A grid-search is performed to find the optimal hyper-parameter settings for each model. 
We use the Adam optimizer~\citep{adam_optimizer} with a learning rate of $0.01$ and decay frequency of 10. 
We set the memory block to 5, maximum epoch to 50, and the dimension of embeddings to 50. 
The gradient is restricted to be at most 40 to prevent gradient explosion; $l2$ regularization is set to 0.001 and the dropout ratio to 0.5. 


\section{Results} 
\label{sec:experiments_results}
\subsection{E1: Comparison with baselines}

Table \ref{tab:results_baseline_babi_dstc2} reports the average turn-level accuracy of \ac{SEntNet} and the baselines on the dialog bAbI and mDSTC2 datasets.%
\footnote{We reimplemented TF-IDF, Q2A, \acs{EntNet} and \ac{SEntNet}. The scores for DQMemNN and HHCN were take from \cite{wu2018end} and \cite{liang2018hierarchical}, respectively; unfortunately, the authors reported performance on only one dataset.} 
\ac{SEntNet} model significantly outperforms all baseline models. 
Furthermore, the \acs{EntNet} model significantly outperforms the simple baselines, i.e., TF-IDF and Q2A, by a large margin. 
Q2A model achieves a surprisingly high turn-level accuracy given the simplicity of the method. 
Furthermore, the large gap between the performance of \ac{SEntNet} and the baseline models suggest that non-trivial relations between the dialogue history and responses are found. 
\ac{SEntNet} is outperformed by HHCN on the mDSTC2 dataset.
This is mainly because HHCN uses a stronger language model (i.e., a word-character RNN) and integrates a NN-based selection for domain knowledge.

\begin{table}[h]
\setlength{\tabcolsep}{4pt}
\centering
\caption{Comparison with baselines on the dialog bAbI and mDSTC2 datasets. The results are the best turn-level accuracy in 10 runs. Bold results indicate a statistically significant improvement over the strongest baseline (paired t-test, $p < 0.01$).}
\label{tab:results_baseline_babi_dstc2}
\begin{tabular}{lrr}
\toprule
{\bfseries Model} &  {\bfseries dialog bAbI} & {\bfseries mDSTC2}\\
\midrule
TF-IDF & 0.040 & 0.030 \\
Q2A & 0.570 & 0.220 \\
\acs{EntNet} & 0.850 & 0.388 \\
DQMemNN & 0.863 & -- \\
HHCN & -- & \textbf{0.661} \\
\midrule
\ac{SEntNet} & \textbf{0.910} & 0.412 \\
\bottomrule
\end{tabular}
\vspace*{-0.5\baselineskip}
\end{table}

\subsection{E2: The effect of embeddings}

\begin{figure}[t]
    \centering
    \begin{tabular}{c}
        \includegraphics[width=0.9\columnwidth]{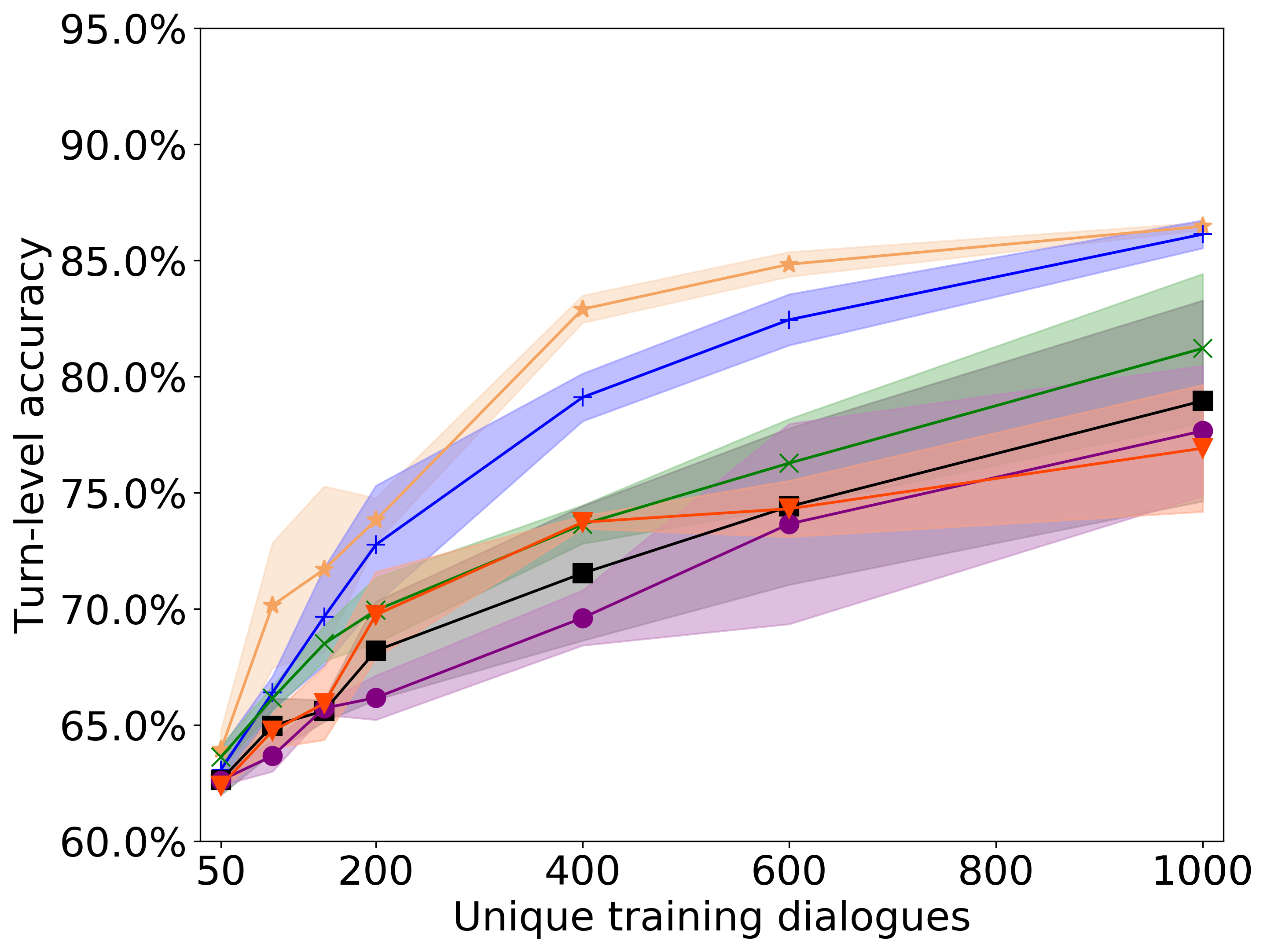} \\
        \small (a) dialog bAbI \\
        \includegraphics[width=0.9\columnwidth]{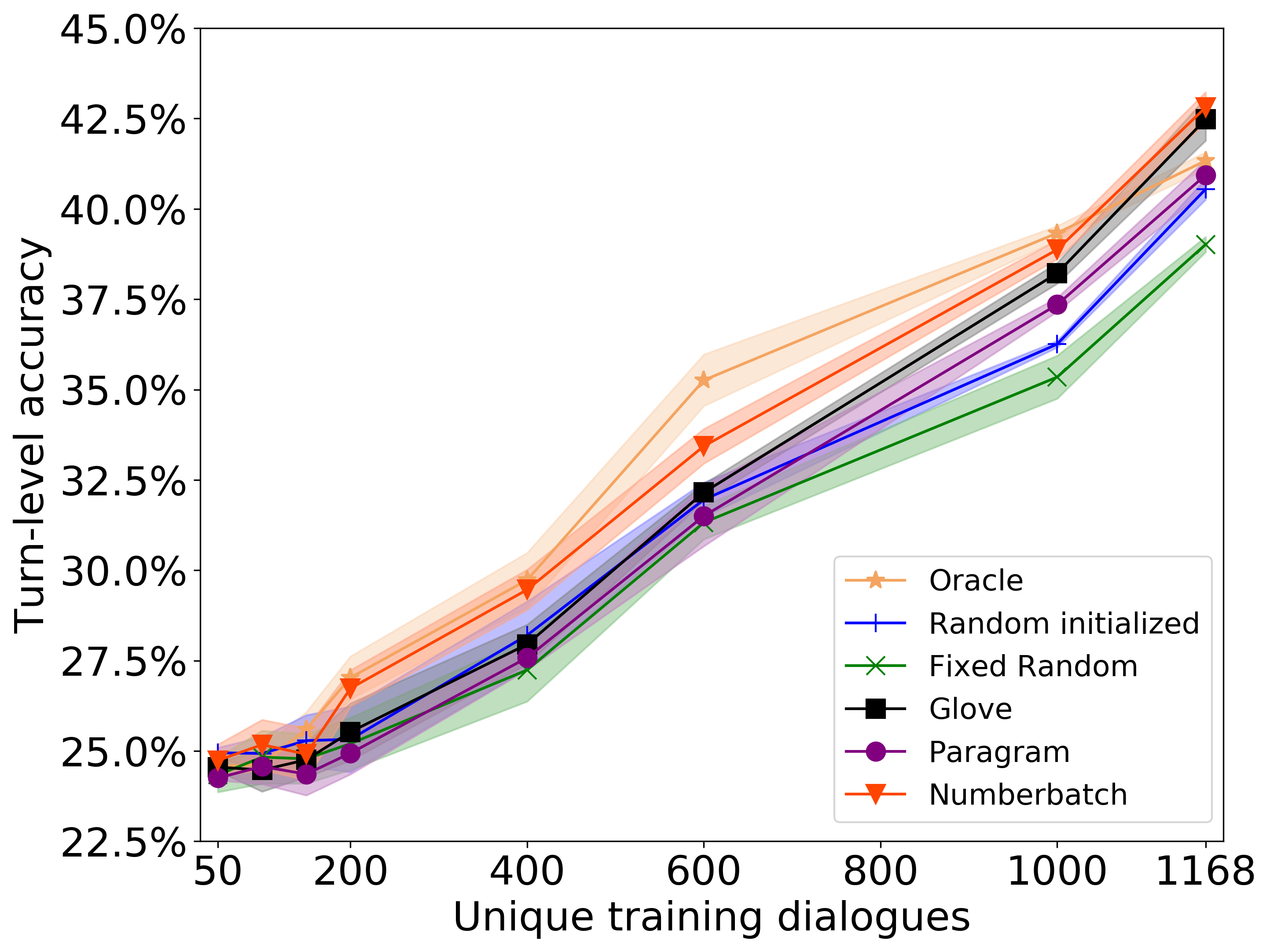} \\
        \small (b) mDSTC2 \\
    \end{tabular}
    \caption{Turn-level accuracy of \ac{SEntNet} for different embedding spaces on both datasets. The accuracy and standard deviation are computed over 10 runs. Please note that the scales on the x-axes and y-axes differ.}
    \label{fig:accuracy_embedding}
        \label{fig:accuracy_embedding_task5}
        \label{fig:accuracy_embedding_task6}
\vspace*{-1\baselineskip}
\end{figure}

To assess the effect of word embeddings, we conduct experiments on the initialization strategies described in Section~\ref{sec:experiments}. 
Figure~\ref{fig:accuracy_embedding} shows the accuracy for different embedding spaces tested on the two datasets.
For both datasets, the \emph{Oracle} strategy outperforms the other strategies when trained on few example dialogues. 
The performance gain diminishes when the number of used training dialogues reaches its maximum.
This indicates that there exists an embedding space that is more effective than randomly initializing embeddings and optimizing during training. 
Additionally, the random strategy outperforms the fixed strategy for both datasets; this indicates that the model can learn useful semantics during training.

For the dialog bAbI dataset, the random strategy has the best performance for any fraction of the dataset. This is not the case for the mDSTC2 dataset, where the optimal embedding strategy depends on the number of available training dialogues. 
The embedding strategies perform equally well when very few example dialogues (less than 200) are available. 
Pre-trained embeddings outperform the randomly initialized embeddings when more dialogues are available. 

Pre-trained embeddings are more useful for the mDSTC2 dataset than the dialog bAbI dataset. 
This is as expected: the overlap in vocabulary between mDSTC2 and the pre-trained embeddings is much larger. 

\subsection{E3: The ability to handle sparse data}

  \begin{figure}[t]
      \centering
      \begin{tabular}{cc}
          \includegraphics[width=0.9\columnwidth]{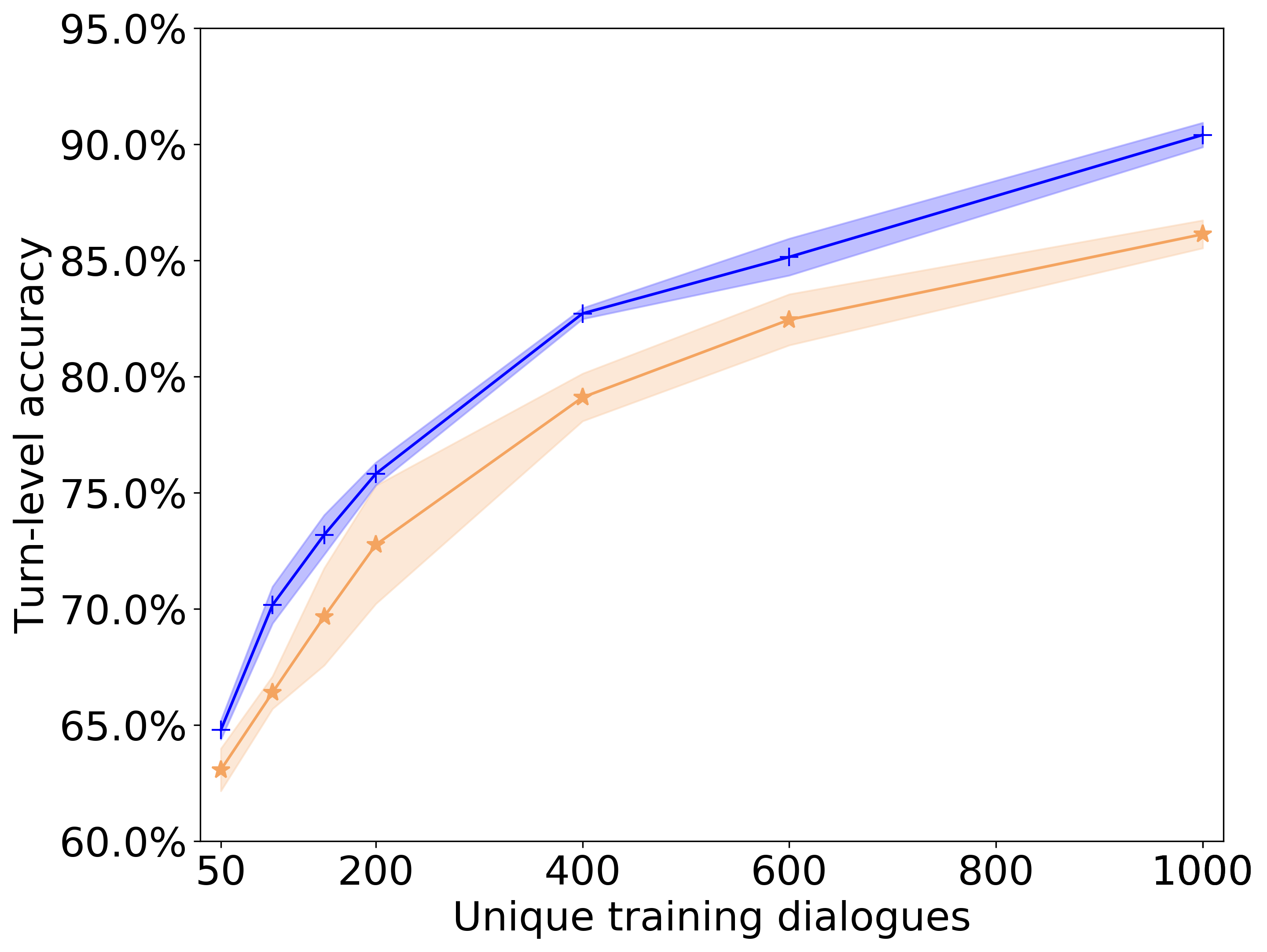} \\
          \small (a) dialogue bAbI \\
          \includegraphics[width=0.9\columnwidth]{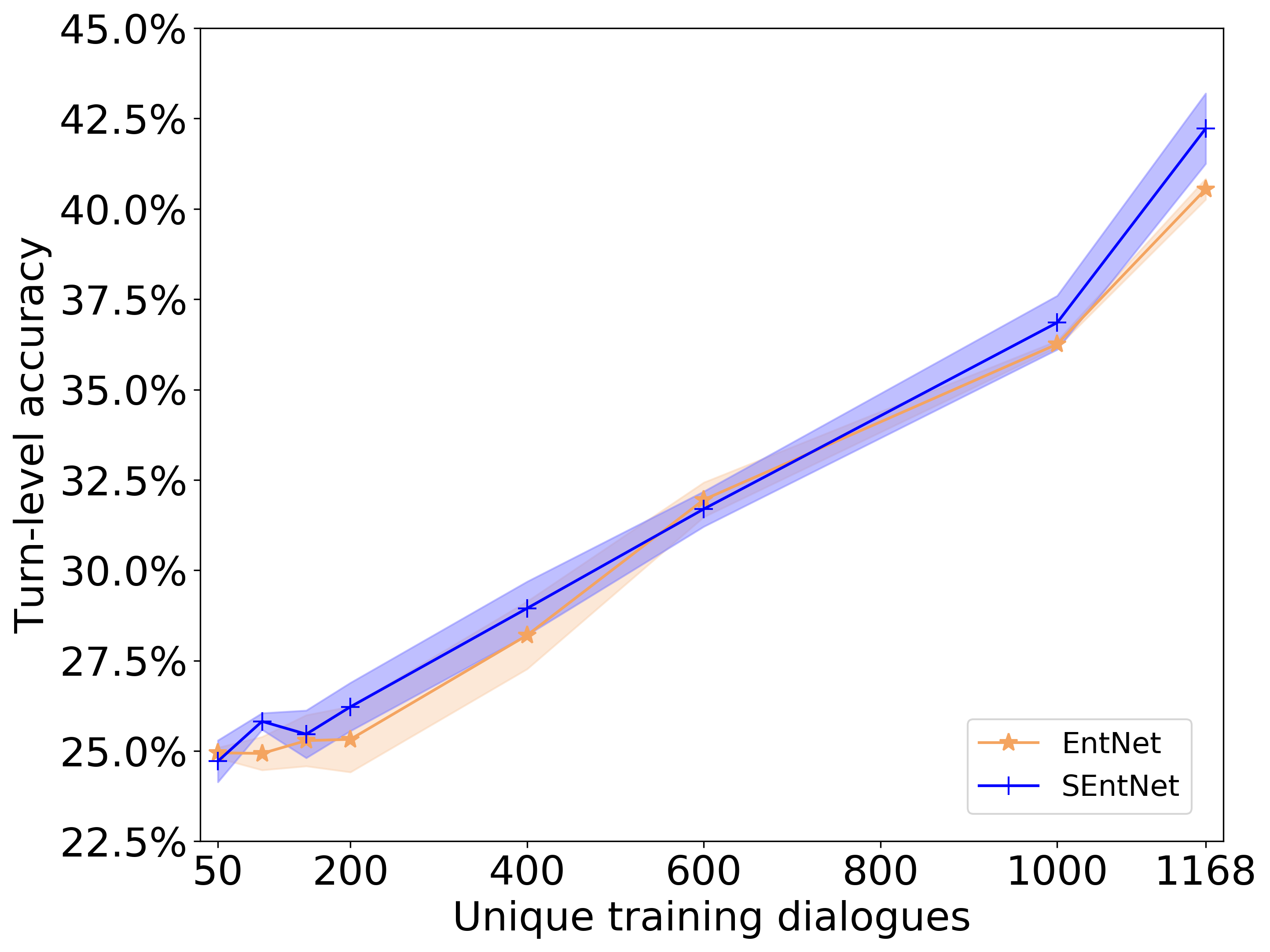} \\
          \small (b) mDSTC2\\
      \end{tabular}
      \caption{Turn-level accuracy of \ac{SEntNet} on both datasets, when trained with different volumes of training dialogues. Please note that the scales on the x-axes and y-axes differ.}
      \label{fig:results_dataset_size}
      \label{fig:results_dataset_size_task5}
      \label{fig:results_dataset_size_dstc2}
   \vspace*{-1\baselineskip}
   \end{figure}

Figure~\ref{fig:results_dataset_size} shows the results of the models we consider when trained on different fractions of the dialog bAbI and mDSTC2 datasets. 
\ac{SEntNet} outperforms \acs{EntNet} for any dialogue size. 
This further supports our hypothesis that \ac{SEntNet} is a better \ac{TDS}. 
The standard deviation of the results seems to show that the difference in performance is significant. 
It shows that \ac{SEntNet} finds a meaningful relationship between the dialogue history and \ac{KB} on the one hand and the response on the other hand, even when few training dialogues are available. 

\subsection{E4: The effect of lexical diversity}
To investigate the effect of lexical diversity, we design a simulation as stated in Section~\ref{sec:experiments}. In this way, we can use \ac{POS} information to generalize concrete words to abstract symbols and get a new synthetic dataset with lower diversity. 

As the results in Table~\ref{fig:results_baseline_babi} show, \ac{SEntNet} outperforms \ac{SEntNet} with \ac{POS} on the dialog bAbI dataset, while \acs{EntNet} with \ac{POS} reaches a similar accuracy as the regular \acs{EntNet}. 
It indicates that \ac{SEntNet} has the potential to handle different degrees of lexical diversity. 
This matters as it may help with adaptation for end-to-end \acp{TDS}.
In the results on the mDSTC2 dataset we do not obtain significant differences; noise in the data may be a crucial factor behind this observation.

\begin{table}[t]
\setlength{\tabcolsep}{4pt}
\centering
\caption{The effect of lexical diversity on \ac{SEntNet} and \acs{EntNet}, on the dialog bAbI and mDSTC2 datasets. The results are the best turn-level accuracy in 10 runs. Bold results indicate a statistically significant improvement over the strongest baseline (paired t-test, $p < 0.01$).} 
\label{fig:results_baseline_babi}
\begin{tabular}{lll}
\toprule
{\bfseries Model} &  {\bfseries dialog bAbI } & {\bfseries mDSTC2}\\
\midrule
\acs{EntNet} & 0.850 & 0.388 \\
\acs{EntNet} + \ac{POS}& 0.850 & 0.398 \\
\midrule
\ac{SEntNet} & \textbf{0.910} & \textbf{0.412} \\
\ac{SEntNet} + \ac{POS} & 0.890 & 0.409 \\
\bottomrule
\end{tabular}
\vspace*{-0.5\baselineskip}
\end{table}


\section{Related Work} 
\label{sec:rel_work}

There are two dominant paradigms for end-to-end dialogue systems: \textit{generation} and \textit{selection}.

\OurParagraph{Generation.}
In this paradigm, a response is generated word by word given the dialogue context. 
Recent improvements have been made by adding an attention mechanism~\citep{vinyals2015neural,li2016persona} or by modeling the hierarchical structure of dialogues~\citep{serban2016building} based on sequential models like \acp{RNN}~\citep{sordoni2015neural}. 
Early attempts~\citep{vinyals2015neural,yao2015attention} to apply these approaches to \acp{TDS} neglect aggregate information from an external \ac{KB}.  
This is problematic when the response contains out-of-dialogue entities.
To address this issue, separate state tracking~\citep{wen2016network} and {soft-lookup} methods~\citep{eric2017key} have been adopted to interact with a \ac{KB}. 

\OurParagraph{Selection.} In this paradigm, a response is selected from a list of candidate responses. 
Important improvements mainly rely on sequential models with attention mechanisms, especially on memory networks.
State-of-the-art models~\citep{end2end_dataset_paper_babi_bordes,liu2017gated,dstc6_memnet_contextnr,wu2018end} are built on top of a MEMN2N architecture~\citep{sukhbaatar2015end} to select a response, where \textit{delexicalization} methods are introduced to enrich an entity with its generic type.
This helps the \ac{TDS} in finding exact matches between entities in the dialogue and the \ac{KB}, and alleviates the out-of-vocabulary problem. 
However, it still explicitly relies on a handcrafted ontology.
Recent models~\citep{end2end_enitity_independent_dstc6,shin2019end} evaluate \acs{EntNet}~\citep{entnet} on DSTC6 dataset~\citep{perezdialog}. 
The models only select the response from a small list of pre-selected candidates. 
Obviously, this does not reflect real-world conversational agents. 
\citet{wu2018end} enhance MEMN2N by dynamic query components originated from \acs{EntNet}, so as to capture the sequential dependencies of dialogue utterances. 
However, their highest performance still relies on delexicalization methods.
\acs{EntNet} can be seen as a group of gated \acp{RNN} with shared parameters, but its hidden states are updated only when they receive new information related to their entities~\citep{entnet}. 
In this way, latent concepts and attributes can be learned by its hidden states~\citep{wu2018end}. 

\smallskip\noindent%
The key distinctions of our work compared to previous efforts are: we introduce a new architecture, \ac{SEntNet}, that adds separate parameterized memory modules to exploit source-specific information.

\section{Conclusion}

We have proposed a dialogue response selection model, \acf{SEntNet}, that is built on top of a memory network architecture and is able to select responses aware of source-specific history for end-to-end \acp{TDS}. 
Experimental results suggest that \ac{SEntNet} consistently outperforms the baselines for end-to-end \ac{TDS}.
Optimizing embeddings while training \ac{SEntNet} is found to be useful for end-to-end task performance.
\ac{SEntNet} is more tolerant of sparse data than baselines and has the potential to handle different degrees of lexical diversity.  

One limitation of \ac{SEntNet} is the increase of learnable parameters with introducing extra memory modules. 
However, the parallel update mechanism design inherited from \acs{EntNet} can offset the use of the computational resources. 
This mechanism makes \ac{SEntNet} scalable to real-world systems that have to deal with even more sources of information.
In future work we plan to apply the source-aware context idea that underlies \ac{SEntNet} to other variant memory networks. 

\OurParagraph{Acknowledgments.}
This research was partially supported by 
Ahold Delhaize,
the Association of Universities in the Netherlands (VSNU),
the China Scholarship Council (CSC),
and
the Innovation Center for Artificial Intelligence (ICAI).
All content represents the opinion of the authors, which is not necessarily shared or endorsed by their respective employers and/or sponsors.

\bibliographystyle{named} 
\bibliography{scai19} 


\end{document}